\documentclass[aps,pra,twocolumn,showpacs,showkeys,superscriptaddress,nobalancelastpage,nofootinbib,floatfix,longbibliography]{revtex4-2}

\usepackage{amsmath,amssymb,amsthm}
\usepackage{graphicx}
\usepackage{physics}
\usepackage{enumitem}
\usepackage{xspace}

\usepackage{empheq} 

\usepackage{orcidlink}

\usepackage{url}
\usepackage{hyperref}
\usepackage{color}
\definecolor{refcolor}{RGB}{160,35,0}
\definecolor{hrefcolor}{RGB}{0,35,190}
\hypersetup{
    colorlinks,
    citecolor=refcolor,
    filecolor=refcolor,
    linkcolor=hrefcolor,
    urlcolor=hrefcolor
}

\def\({\left(}
\def\){\right)}
\def\[{\left[}
\def\]{\right]}

\newcommand{\pobs}[1]{#1}
\newcommand{\obs}[1]{\mathsf{\pobs{#1}}}

\newcommand{\R}{\mathbb{R}}

\newcommand{\WdWSpaceM}{{\mkern-6mu}}
\newcommand{\WdWSpaceMM}{{\mkern-9mu}}
\newcommand{\WdWSpaceR}{{\mkern-8mu}}

\newcommand{\kett}[1]{\ket{\WdWSpaceM\ket{#1}\WdWSpaceR}}

\newcommand{\brakett}[2]{\left.\bra{#1}{\WdWSpaceMM}\ket{#2}{\WdWSpaceR}\right\rangle}

\newcommand{\x}{\mathbf{x}}

\newcommand{\q}{\boldsymbol{q}}
\newcommand{\M}{\mathbf{M}}

\newcommand{\ie}{\textit{i.e.}\xspace}

\newcommand{\etc}{\textit{etc}\xspace}

\newcommand{\sref}[1]{\S\ref{#1}}

\newcommand{\image}[3]{
\begin{figure}[!ht]
\centering
\includegraphics[width=#2\textwidth]{#1}
\caption{\small{\label{#1}#3}}
\end{figure}
}

\newtheorem{question}{Question}

\theoremstyle{remark}

\newtheoremstyle{nospace} 
  {3pt}                   
  {3pt}                   
  {\itshape}              
  {}                      
  {\bfseries}             
  {.}                     
  {.5em}                  
  {\thmname{#1}\thmnumber{#2}\thmnote{ (#3)}} 

\theoremstyle{nospace}

\newtheorem{defCustom}{}
\makeatletter
\newcommand{\setdefCustomtag}[1]{
  \let\oldthedefCustom\thedefCustom
  \renewcommand{\thedefCustom}{{\normalfont\textbf{#1}}}
  \g@addto@macro\enddefCustom{
    \global\let\thedefCustom\oldthedefCustom}
  }
\makeatother

\allowdisplaybreaks

\begin{document}

\title{Emergent cosmology and gravity from quantum time?}
\author{\orcidlink{0000-0002-2765-1562}\ O.C. Stoica}
\affiliation{Dept. Th. Physics, NIPNE-HH, Bucharest, Romania. \href{mailto:cristi.stoica@theory.nipne.ro}{cristi.stoica@theory.nipne.ro},  \href{mailto:holotronix@gmail.com}{holotronix@gmail.com}}



\begin{abstract}
Macroscopic observables allow the recovery of intrinsic dynamics from stationary quantum states. I show that, by interpreting the squared amplitude as the probability density for each definite value of intrinsic time, a curvature emerges in the time direction. For example, from the perspective of intrinsic quantum time, the Friedmann-Lema\^itre-Robertson-Walker cosmological model emerges from spherically symmetric stationary solutions in four-dimensional Euclidean space, without presupposing gravity. If there is no unique direction of time, curvature emerges in all spacetime dimensions, without presupposing gravity, from the variable amplitude of the stationary wavefunction alone. This opens a new possibility that general relativity or some modification of it emerges from intrinsic time observables.
\end{abstract}

\maketitle

\section{Introduction}
\label{s:intro}

In canonical quantum gravity and quantum cosmology, time evolution seems to vanish, resulting in the Wheeler-DeWitt constraint equation~\cite{Dewitt1967QuantumTheoryOfGravityI_TheCanonicalTheory,Dewitt1967QuantumTheoryOfGravityII_TheManifestlyCovariantTheory,Halliwell1991IntroductoryLecturesOnQuantumCosmology,Wiltshire1996AnIntroductionToQuantumCosmology,Kiefer2012QuantumGravity,KieferSandhofer2022QuantumCosmology},
\begin{equation}
\label{eq:WDW}
\obs{H}\kett{\Psi}=0.
\end{equation}

This condition seems to be a characteristic of any quantum theory that includes a long-range gravitational coupling~\cite{PageWootters1983EvolutionWithoutEvolution}, for the same reason why the long range of electromagnetic interactions requires the existence of a charge superselection rule~\cite{WickWightmanWigner1952TheIntrinsicParityOfElementaryParticles}.

But this timelessness is in fact much more general, being an effect of intrinsic time observables~\cite{Stoica2026QuantumTimeWithoutNegativeEnergy}. If the Hamiltonian is bounded from below, exact time observables monotonically correlated with the Schr\"odinger time $t$ seem to be forbidden~\cite{UnruhWald1989TimeAndTheInterpretationOfCanonicalQuantumGravity}. In fact, any definite irreversible observable change seems to be forbidden~\cite{HegerfeldtRuijsenaars1980RemarksOnCausalityLocalizationAndSpreadingOfWavePackets}. But the accuracy of our empirical data, or even the very existence of classical and quantum computing~\cite{Stoica2025CanWeAccuratelyReadWriteQuantumData} seem to confirm the existence of irreversible changes.
This invites us to acknowledge the possibility of exact time observables, even if they are not monotonically correlated with the Schr\"odinger parameter $t$.
Demoting the Schr\"odinger parameter $t$ from the traditional role of time and using pointer states as time observables lead to the same equation~\eqref{eq:WDW}, even without assuming gravity~\cite{Stoica2026QuantumTimeWithoutNegativeEnergy}.

In the proposal from~\cite{Stoica2026QuantumTimeWithoutNegativeEnergy} and other approaches like the Page-Wootters proposal~\cite{PageWootters1983EvolutionWithoutEvolution,Wootters1984TimeReplacedByQuantumCorrelations,Page1986DensityMatrixOfTheUniverse,Page1989TimeAsAnInaccessibleObservable,Page1994ClockTimeAndEntropy,GiovannettiLloydMaccone2015QuantumTime,LoveridgeMiyadera2019RelativeQuantumTime,MacconeSacha2020QuantumMeasurementsOfTime,CastroruizGiacominiBelenchiaBrukner2020QuantumClocksAndTemporalLocalisabilityOfEventsInThePresenceOfGravitatingQuantumSystems,FotiEtAl2021TimeAndClassicalEqOfMotionFromQEntanglementViaPageWoottersGeneralizedCoherentStates,HohnSmithLock2021TrinityOfRelationalQuantumDynamics,Gambini2022SolutionProblemOfTimeQuantumGravityAlsoTimeArrivalProblemQM,AltaieBeigeHodgson2022TimeAndQuantumClocksAReviewOfRecentDevelopments,Adlam2022WatchingTheClocksInterpretingPageWoottersFormalismAndInternalQRF,Rijavec2022HeisenbergPictureEvolutionWithoutEvolution,Rijavec2023robustnessOfThePageWoottersConstructionAcrossDifferentPicturesStatesOfTheUniverseAndSystemClockInteractions,SuleymanovCohen2023QuantumFramesOfReferenceAndRelationalFlowOfTime,RidleyCohen2025TwoTimesOrNone}, we can treat time observables in the following way.
We read time from macroscopic systems like clocks, planetary systems, the cosmic background radiation, \etc. Thus, we use pointer states, and the properties we read are macroscopic. Let $\widehat{\M}=(\obs{M}_1,\obs{M}_2,\ldots)$ be a set of macroscopic commuting observables large enough so that all collective macroscopic observables can be obtained as functions of $\widehat{\M}$. Our time observable is among them, so we can choose $\widehat{\M}$ as commuting observables that also commute with $\obs{T}$.
We extend $\{\obs{T}\}\cup\widehat{\M}$ to a complete set of commuting observables $\{\obs{T}\}\cup\widehat{\M}\cup\widehat{\q}$, where $\widehat{\q}=(\widehat{q}_1,\widehat{q}_2,\ldots)$. 
The possible eigenvalues $(\tau,\M,\q)$ of $\obs{T}$, $\widehat{\M}$, and $\widehat{\q}$, corresponding to a common eigenbasis of the operators $\obs{T}$, $\widehat{\M}$, and $\widehat{\q}$, form a configuration space for the wavefunction of the universe $\psi$, that is,
\begin{equation}
\label{eq:t-wavefunction}
\psi(\tau,\M,\q;t)=\braket{\tau,\M,\q}{\psi(t)}.
\end{equation}

The stationary state vector $\kett{\Psi}$ as in equation~\eqref{eq:WDW} yields a timeless wavefunction~\cite{Stoica2026QuantumTimeWithoutNegativeEnergy}
\begin{equation}
\label{eq:tau-wavefunction-from-t}
\Psi(\tau,\M,\q)=\brakett{\tau,\M,\q}{\Psi}=\int_{\R}\psi(\tau,\M,\q;t)\dd t.
\end{equation}

The regularized integral can be nonzero only if $0$ is in the spectrum of the Hamiltonian. In this case, one obtains that $\obs{H}\kett{\Psi}=0$. More details about this representation can be found in~\cite{Stoica2026QuantumTimeWithoutNegativeEnergy}.
This representation works equally well with the Page-Wootters proposal and with the other proposals.

The squared amplitude 
\begin{equation}
\label{eq:prob-tau}
P(\tau):=\int\abs{\Psi(\tau,\M,\q)}^2\dd\M\dd\q
\end{equation}
can be interpreted as the (unnormalized) probability density that the intrinsic time is $\tau$. We can use an unnormalized probability density, if we compare the relative probabilities only locally.
It may happen that $P(\tau)$ is not uniform with respect to $\tau$.
For example, in a Page-Wootters system, the squared amplitude may assign nonuniform weights to the time reading $\tau$, that is,
\begin{equation}
\label{eq:PW-weight}
\kett{\Psi}=\int_{\R}c(\tau)\ket{\tau}_c\ket{\psi(\tau)}_r\dd\tau
\end{equation}
where $\abs{c(\tau)}^2$ is not independent of $\tau$.
While this does not affect the conditional expectation value~\cite{PageWootters1983EvolutionWithoutEvolution,Wootters1984TimeReplacedByQuantumCorrelations,Page1986DensityMatrixOfTheUniverse}, it affects the rate of flow of time, by changing the probability that an observer finds herself in one time interval rather than another.
In Section~\ref{s:emergent-cosmology} we will see more concrete examples.

When possible, we can reparametrize monotonically the spectrum of $\obs{T}$, obtaining another time observable with the same generalized eigenvectors, but the spectrum redistributed so that $P(\tau)$ is uniform with respect to $\tau$, roughly,
\begin{equation}
\label{eq:prob-tau-uniform}
P(\tau)=\text{const.}
\end{equation}

This uniformizes the self-location probability with respect to the intrinsic time.
It was shown that this condition also leads to an effective intrinsic Schr\"odinger equation with respect to $\tau$~\cite{Stoica2026QuantumTimeWithoutNegativeEnergy}.

In Section~\ref{s:emergent-cosmology}, we will see that, in the case of a spherically symmetric stationary wavefunction in four-dimensional Euclidean space, this uniformization of the time observable leads to the closed Friedmann-Lema\^itre-Robertson-Walker (FLRW) cosmological model.
We can choose the four-dimensional potential to reproduce any expanding homogeneous and isotropic cosmology, and several examples are given, including the $\Lambda$CDM model.

This amounts to acquiring curvature on a flat space just from the quantum time observable, without including gravity explicitly.
One may say that gravity is implicit in the potential $V(r)$, but even for $V(r)=0$ we obtain a nonvanishing curvature (Subsection~\ref{s:emergent-cosmology-free}).
Therefore, the intrinsic time observable leads to both a Wheeler-DeWitt-type equation~\cite{Stoica2026QuantumTimeWithoutNegativeEnergy} and FLRW cosmology, which includes curvature in the time direction, without presupposing gravity. Can it also generate a spacetime curvature like in general relativity?
I will explore this possibility in Section~\ref{s:emergent-curvature}, using four independent time observables.
Is it possible that general relativity, some additional curvature effect, or even modified gravity~\cite{Milgrom1983AModificationOfTheNewtonianDynamicsAsAPossibleAlternativeToTheHiddenMassHypothesis,Bekenstein2004RelativisticGravitationTheoryForTheMONDParadigm,FamaeyMcGaugh2012ModifiedNewtonianDynamicsMONDObservationalPhenomenologyAndRelativisticExtensions} emerge from quantum theory without presupposing gravity?

A central question that such a program needs to answer is raised and discussed in Section~\ref{s:uniformity-problem}: since the rate of time flow, and therefore the curvature, are observable, the amplitudes should be encoded in the intrinsic records.

\section{Toy emergent cosmologies from intrinsic time}
\label{s:emergent-cosmology}

According to the \emph{cosmological principle}, the large-scale distribution of matter in the universe is isotropic and homogeneous.
Under this assumption, classical general relativity leads to the Friedmann-Lema\^itre-Robertson-Walker cosmological model, whose metric tensor is 
\begin{equation}
\label{eq:FLRW-metric-a}
\dd s^2=-c^2 \dd t^2+a^2(t)\,\dd\Sigma^2.
\end{equation}

Here $a(t)$ is a dimensionless scale parameter, and $\dd\Sigma$ is the line element of a $3$-dimensional space of uniform curvature, which can be the 3-sphere $S^3$ ($k=1$), the Euclidean space ($k=0$), or the hyperbolic space ($k=-1$).

In cosmology, one often relies on the simplifying assumption that the matter filling the universe is a perfect fluid with energy-momentum tensor
\begin{equation}
\label{eq:energy-momentum-perfect-fluid}
T_{\mu\nu}=\(\rho c^2+p\)u_\mu u_\nu+p g_{\mu\nu},
\end{equation}
where $\rho$ is the density and $p$ is the pressure.

The Einstein field equations
\begin{equation}
\label{eq:einstein}
G_{\mu\nu}+\Lambda g_{\mu\nu}=\frac{8\pi G}{c^4} T_{\mu\nu}
\end{equation}
become, for the metric~\eqref{eq:FLRW-metric-a} and the energy-momentum tensor~\eqref{eq:energy-momentum-perfect-fluid}, the \emph{Friedmann equations}
\begin{subequations}
\label{eq:friedmann-scale}
\begin{empheq}[left={\empheqlbrace}]{align}
\frac{\dot a^2}{a^2}&=\phantom{-}\frac{8\pi G}{3}\rho-\frac{k c^2}{a^2}+\frac{\Lambda c^2}{3} \label{eq:friedmann-scale-hubble}\\
\frac{\ddot a}{a}&=-\frac{4\pi G}{3}\(\rho+3\frac{p}{c^2}\)+\frac{\Lambda c^2}{3} \label{eq:friedmann-scale-acc}
\end{empheq}
\end{subequations}
where $\dot a=\dv*{a}{t}$ and $\ddot a=\dv*[2]{a}{t}$.
From them one can deduce the explicit form of energy-momentum conservation,
\begin{equation}
\label{eq:friedmann-rho-a}
\dot\rho =-3\frac{\dot a}{a}\(\rho+\frac{p}{c^2}\).
\end{equation}

In quantum cosmology, one usually starts with a classical cosmological model and then quantizes it. One obtains a universal wavefunction on a reduced configuration space called \emph{minisuperspace}, which consists of a small number of independent degrees of freedom such as the scale parameter $a$ and one or more homogeneous scalar fields. One solves the Wheeler-DeWitt equation by applying the WKB approximation to get solutions of the form $\Psi(a)\sim A(a)e^{-i S(a)}$.
To get a classical approximate solution, one interprets the phase $S(a)$ as the classical action, then the WKB approximation gives the constrained, time-independent Hamilton-Jacobi equation. From this, one interprets the derivative $\dv*{S}{a}$ as classical momentum, obtaining semiclassical solutions of the Wheeler-DeWitt equation.
In the classical limit one obtains a classical cosmological model~\cite{HartleHawking1983WaveFunctionOfTheUniverse,Vilenkin1984QuantumCreationOfUniverses,Halliwell1991IntroductoryLecturesOnQuantumCosmology,Wiltshire1996AnIntroductionToQuantumCosmology,Linde1998QuantumCreationOfAnOpenInflationaryUniverse,AshtekarSingh2011LoopQuantumCosmologyAStatusReport,Kiefer2012QuantumGravity,KieferSandhofer2022QuantumCosmology}.

Here, as an application of the intrinsic time observable, I will simply start with a spherically symmetric Hamiltonian in a four-dimensional Euclidean space and a scalar stationary solution, pick a uniform intrinsic time observable invariant under the spherical symmetry, and show that it gives an intrinsic view that leads immediately to a closed FLRW cosmological model of an expanding universe, even though no gravity exists in the original Hamiltonian. In particular, we will obtain the $\Lambda$CDM model, the current Standard Model of modern cosmology.
It may be premature to regard what follows as \emph{the} quantum cosmology. I propose it instead as a proof of concept that curvature can emerge as an effect of intrinsic time.

\subsection{Spherically symmetric stationary solutions}
\label{s:WDW-spherical}

We consider a four-dimensional Schr\"odinger equation with a spherical potential $V(r)$,
\begin{equation}
\label{eq:hamiltonian-spherical-four}
\obs{H}=-\frac{\hbar^2}{2m}\laplacian+V(r)
\end{equation}
where
\begin{equation}
\label{eq:Laplacian}
\laplacian:=\pdv[2]{x_1}+\pdv[2]{x_2}+\pdv[2]{x_3}+\pdv[2]{x_4}
\end{equation}
is the Laplacian in four dimensions and
\begin{equation}
\label{eq:r}
r:=\abs{\x}=\sqrt{x_1^2+x_2^2+x_3^2+x_4^2}.
\end{equation}

The potential is assumed to allow the existence of a zero-energy solution.
Then, the Schr\"odinger equation with the parameter $t$ is
\begin{equation}
\label{eq:schrod-four}
i\hbar\pdv{\psi}{t}=\[-\frac{\hbar^2}{2m}\laplacian+V(r)\]\psi
\end{equation}

We work in spherical coordinates, with the radial coordinate $r$.
The Laplacian's action on a spherically symmetric solution $\psi(r,t)$ is
\begin{equation}
\label{eq:laplacian}
\laplacian\psi=\pdv[2]{\psi}{r}+\frac{3}{r}\pdv{\psi}{r},
\end{equation}
so the Schr\"odinger equation becomes
\begin{equation}
\label{eq:schrod-four-spherical}
i\hbar\pdv{\psi}{t}=-\frac{\hbar^2}{2m}\[\pdv[2]{\psi}{r}+\frac{3}{r}\pdv{\psi}{r}\]+V(r)\psi.
\end{equation}

Due to equation~\eqref{eq:WDW}, we are interested in zero-energy stationary solutions $\kett{\Psi}$.
Due to the spherical symmetry, we will choose the intrinsic time observable $\tau$ to depend on $r$, $\tau=\tau(r)$. But, while the density $\abs{\Psi(r)}^2$ is, in general, unevenly distributed as a function of $r$, we choose $\tau$ to be distributed uniformly.

Then, we will see that $r(\tau)$ expands as $\tau$ increases, so that we obtain the Friedmann-Lema\^itre-Robertson-Walker cosmological model of an expanding universe.
Since the function $r(\tau)$ is not linear, the resulting universe is expanding and curved in the intrinsic time dimension without assuming gravity, and this is true even if the potential $V(r)=0$.
We can choose $V(r)$ so that we obtain the $\Lambda$CDM model, which is the current Standard Model of cosmology.

Due to spherical symmetry, the zero-energy stationary solution $\Psi$ depends only on $r$, and so it satisfies

\begin{equation}
\label{eq:radial-eigenvalue}
-\frac{\hbar^2}{2m}\(\dv[2]{\Psi}{r}+\frac{3}{r}\dv{\Psi}{r}\)+V(r)\Psi(r)=0.
\end{equation}

This becomes the radial eigenvalue equation
\begin{equation}
\label{eq:radial-eigenvalue-eq}
\Psi''(r)+\frac{3}{r}\Psi'(r)-\frac{2m}{\hbar^2}V(r)\Psi(r)=0.
\end{equation}

To simplify it, let us define the reduced radial function
\begin{equation}
u(r)=r^{\frac{3}{2}}\Psi(r).
\end{equation}

Then,
\begin{equation}
\label{eq:Psi-of-u}
\Psi(r)=r^{-\frac{3}{2}}u(r).
\end{equation}

By straight computation,
\begin{equation}
\Psi'(r)=r^{-\frac{3}{2}}u'(r)-\frac{3}{2}r^{-\frac{5}{2}}u(r)
\end{equation}
and
\begin{equation}
\Psi''(r)=r^{-\frac{3}{2}}u''(r)-3r^{-\frac{5}{2}}u'(r)
+5r^{-\frac{7}{2}}u(r).
\end{equation}

This gives
\begin{equation}
\Psi''(r)+\frac{3}{r}\Psi'(r)=r^{-\frac{3}{2}}\(u''(r)-\frac{3}{4r^2}u(r)\).
\end{equation}

By substituting this into~\eqref{eq:radial-eigenvalue} we get
\begin{equation}
\label{eq:reduced-schrod}
-\frac{\hbar^2}{2m}\(u''(r)-\frac{3}{4r^2}u(r)\)+V(r)u(r)=0.
\end{equation}

Therefore, $u(r)$ satisfies the one-dimensional Schr\"odinger-type equation
\begin{equation}
\label{eq:schrod-one-dim-u}
-\frac{\hbar^2}{2m}u''(r)+V_{\mathrm{eff}}(r)u(r)=0
\end{equation}
where $V_{\mathrm{eff}}(r)$ is an effective radial potential
\begin{equation}
\label{eq:potential-effective}
V_{\mathrm{eff}}(r):=V(r)+\frac{\hbar^2}{2m}\frac{3}{4r^2}.
\end{equation}

Then, the solutions to equation~\eqref{eq:radial-eigenvalue} have the form~\eqref{eq:Psi-of-u}, where $u$ satisfies equation~\eqref{eq:schrod-one-dim-u}.

The four-dimensional volume element is
\begin{equation}
\dd^4x=r^3\,\dd r\,\dd\Omega_3,
\end{equation}
where $\dd\Omega_3$ is the dimensionless area element on the unit $3$-sphere $S^3$,
\begin{equation}
\int_{S^3}\dd\Omega_3=\operatorname{Area}(S^3)=2\pi^2.
\end{equation}

If $\Psi$ is normalized so that
\begin{equation}
\label{eq:Psi-four-normalized}
\int_{\mathbb{R}^4}\abs{\Psi(\x)}^2\,\dd^4x=1,
\end{equation}
then, in radial variables
\begin{equation}
\label{eq:Psi-rad-normalized}
\int_0^\infty \abs{\Psi(r)}^2 r^3\,\dd r=\frac{1}{2\pi^2}.
\end{equation}

Since $u(r)=r^{3/2}\Psi(r)$, this implies
\begin{equation}
\label{eq:u-rad-normalized}
\int_0^\infty \abs{u(r)}^2\,\dd r=\frac{1}{2\pi^2}.
\end{equation}

The densities or differential forms $\abs{\Psi(\x)}^2\,\dd^4x$ from~\eqref{eq:Psi-four-normalized}, $\abs{\Psi(r)}^2 r^3\,\dd r$ from~\eqref{eq:Psi-rad-normalized}, and $\abs{u(r)}^2\,\dd r$ from~\eqref{eq:u-rad-normalized} are dimensionless.
This implies that, if $L$ is a unit of length, in this formulation,
\begin{itemize}
	\item the unit of $\Psi(\x)$ and $\Psi(r)$ is $L^{-2}$, and
	\item the unit of $u(r)$ is $L^{-1/2}$.
\end{itemize}

\subsection{The uniform intrinsic time observable}
\label{s:intrinsic-time}

Now we define a spherically symmetric intrinsic time observable $\obs{T}$ so that the squared amplitude of $\Psi$ is uniformly distributed with respect to its eigenvalues $\tau$.
The integral of the squared amplitude over the shell $S_r^3$ is
\begin{equation}
\label{eq:I-of-r}
I(r)=\int_{S_r^3}\abs{\Psi(r)}^2\,dA_r=2\pi^2\abs{\Psi(r)}^2r^3,
\end{equation}
because $dA_r=r^3\,\dd\Omega_3$ and $\Psi$ is constant on each shell, being spherically symmetric.
Its dimension is $L^{-1}$.

In terms of $u$, this becomes
\begin{equation}
I(r)=2\pi^2\abs{u(r)}^2.
\end{equation}

In a thin radial shell layer between $r$ and $r+\dd r$, the total squared amplitude is the dimensionless quantity
\begin{equation}
\dd P=I(r)\,\dd r.
\end{equation}

We define a uniform intrinsic time $\tau$ as the accumulated shell measure
\begin{equation}
\label{eq:tau-of-r}
\tau(r)=T\int_0^r I(s)\,ds,
\end{equation}
where $T>0$ is a fixed scale factor that can be chosen, when possible, so that the restriction to the level hypersurface $\tau=\text{const.}$, $\ket{\psi(\tau)}=\obs{P}_{\tau}\kett{\Psi}$, is normalized.
This normalization condition sets the unit of time $T$.
If $\kett{\Psi}$ is normalized, $\tau(r)$ is the cumulative distribution function multiplied by $T$ and the inverse $r(\tau)$ is the quantile function divided by $T$. But we will not require $\kett{\Psi}$ to be normalized, because only $\ket{\psi(\tau)}$ has to be normalized.

We get
\begin{equation}
\label{eq:dtau-of-dr}
\dd\tau=T I(r)\,\dd r.
\end{equation}
Conversely,
\begin{equation}
\label{eq:dr-of-dtau}
\dd r=\frac{\dd\tau}{T I(r)}.
\end{equation}

The intrinsic time coordinate $\tau(r)$ is monotone in $r$.
If it is strictly monotone it is invertible, and its inverse is $r(\tau)$.
In the shell layer between $r(\tau)$ and $r(\tau+\dd\tau)$, the amount of probability is
\begin{equation}
\label{eq:dP-of-p-dr}
\dd P=I(r(\tau))\dd r.
\end{equation}

From equations~\eqref{eq:dP-of-p-dr} and \eqref{eq:dtau-of-dr} we can check that indeed $\tau$ uniformizes the probability density,
\begin{equation}
\dv{P}{\tau}=\frac{1}{T}.
\end{equation}

This implies that $\tau$ is a uniform coordinate, and it can be used to define the intrinsic time operator
\begin{equation}
\label{eq:intrinsic-tau-spherical}
\obs{T}:=\tau(\widehat{r})=\int_0^\infty \tau(r)\dyad{r}\,\dd r,
\end{equation}
where $\widehat{r}$ is the radial position observable.
Then, the generalized eigenvectors of $\obs{T}$ that satisfy the uniformization condition are
\begin{equation}
\label{eq:intrinsic-tau-eigenvectors}
\ket{\tau}=\frac{1}{\sqrt{\abs{\dv{\tau}{r}\(r(\tau)\)}}}\ket{r(\tau)}=\frac{1}{\sqrt{\abs{T I\(r(\tau)\)}}}\ket{r(\tau)}.
\end{equation}

Due to the scaling property of the Dirac distribution, we can choose the scale factor $T$ so that, for all $\tau,\tau'$, 
\begin{equation}
\label{eq:tau-basis}
\braket{\tau}{\tau'}_c=\delta(\tau-\tau').
\end{equation}

\subsection{A toy emergent FLRW cosmology}
\label{s:emergent-cosmology-basics}

Even if the wavefunction $\Psi$ is defined on a Euclidean space, since the spherical symmetry allowed a preferred foliation by the intrinsic time, we can naturally define a Lorentzian metric by the line element
\begin{equation}
\label{eq:FLRW-metric-r}
\dd s^2=-c^2 \dd\tau^2+\frac{r^2(\tau)}{L^2}\,\dd\Sigma^2,
\end{equation}
obtaining a FLRW model like~\eqref{eq:FLRW-metric-a}.
There is a binary choice here, we could equally define an Euclidean metric $\dd s^2=c^2 \dd\tau^2+\frac{r^2(\tau)}{L^2}\,\dd\Sigma^2$.

Since $r$ has the dimension of length, it is related to the dimensionless scale factor $a$ by
\begin{equation}
\label{eq:r-of-a}
r(\tau)=a(\tau)L.
\end{equation}

Therefore, with the intrinsic time $\tau$, the stationary solution resembles the classical FLRW cosmological model of a closed expanding universe, $\dd s^2=-c^2 \dd\tau^2+a^2(\tau)\,\dd\Sigma^2$ with $k=1$.

This Lorentzian metric is singular at $r=0$ ($\tau=0$), corresponding to the Big Bang, even though the original Euclidean metric $\dd r^2+r^2(\tau)\,\dd\Sigma^2/L^2$ is not singular.

Using the dot notation for the derivatives with respect to $\tau$, the Friedmann equations~\eqref{eq:friedmann-scale} become
\begin{equation}
\label{eq:friedmann-r}
\left\{
\begin{aligned}
\frac{\dot r^2}{r^2}&=\phantom{-}\frac{8\pi G}{3}\rho-\frac{c^2}{r^2}+\frac{\Lambda c^2}{3}\\
\frac{\ddot r}{r}&=-\frac{4\pi G}{3}\(\rho+3\frac{p}{c^2}\)+\frac{\Lambda c^2}{3},\\
\end{aligned}
\right.
\end{equation}
and the explicit form~\eqref{eq:friedmann-rho-a} of energy-momentum conservation becomes
\begin{equation}
\label{eq:friedmann-rho-r}
\dot\rho =-3\frac{\dot r}{r}\(\rho+\frac{p}{c^2}\).
\end{equation}

To solve for $\rho$ and $p$, we need to know the \emph{Hubble parameter} $\dot r/r$ and $\ddot r/r$.

From equation~\eqref{eq:dr-of-dtau},
\begin{equation}
\label{eq:dot-r}
\dot r=\dv{r}{\tau}=\frac{1}{T I(r)}.
\end{equation}

From this and equation~\eqref{eq:tau-of-r}, the Hubble parameter is
\begin{equation}
\label{eq:hubble}
\frac{\dot r}{r}=\frac{1}{
2\pi^2 T\abs{\Psi(r)}^2 r^4}
\end{equation}
or, in terms of the reduced radial function $u(r)$,
\begin{equation}
\label{eq:hubble-u}
\frac{\dot r}{r}=\frac{1}{
2\pi^2 T\abs{u(r)}^2 r}.
\end{equation}

By differentiating equation~\eqref{eq:dot-r} again and using the chain rule $\dot I(r)=I'(r)\dot r$, we get
\begin{equation}
\label{eq:r-ddot}
\ddot r=-\frac{1}{T^2}\frac{I'(r)}{I(r)^3}.
\end{equation}

We compute $I'(r)$ using equation~\eqref{eq:I-of-r},
\begin{equation}
\label{eq:I-of-r-prime}
I'(r)=2\pi^2\(r^3\dv{r}\abs{\Psi(r)}^2+3r^2\abs{\Psi(r)}^2\).
\end{equation}

Then, from equations~\eqref{eq:r-ddot} and~\eqref{eq:I-of-r-prime} we obtain
\begin{equation}
\frac{\ddot r}{r}=-\frac{1}{T^2}\frac{r^3\dv{r}\abs{\Psi(r)}^2+3r^2\abs{\Psi(r)}^2}{
4\pi^4\abs{\Psi(r)}^6 r^{10}}.
\end{equation}

Or, in terms of the reduced radial function $u$,
\begin{equation}
\frac{\ddot r}{r}=-\frac{1}{T^2}\frac{\dv{r}\abs{u(r)}^2}{4\pi^4 \abs{u(r)}^6 r}.
\end{equation}

To solve for $\rho$ and $p$, we plug the expressions of $\dot r/r$ and $\ddot r/r$ in the Friedmann equations~\eqref{eq:friedmann-r}.

\subsection{Example: The harmonic oscillator potential}
\label{s:emergent-cosmology-harmonic}

For a simple example, consider the isotropic four-dimensional harmonic oscillator with shifted potential
\begin{equation}
V(r)=\frac{1}{2}m\omega^2 r^2 - 2\hbar\omega.
\end{equation}

Let us denote the oscillator length by
\begin{equation}
\mathfrak{a}=\sqrt{\frac{\hbar}{m\omega}}.
\end{equation}

The zero-energy stationary four-dimensional normalized solution is
\begin{equation}
\Psi(r)=\frac{1}{\mathfrak{a}^2\pi}e^{-r^2/(2\mathfrak{a}^2)}.
\end{equation}

The integral of the squared amplitude over the shell $S_r^3$ is
\begin{equation}
\label{eq:I-of-r-harm}
I(r)=\frac{2}{\mathfrak{a}^4}r^3e^{-r^2/\mathfrak{a}^2}.
\end{equation}

The intrinsic time coordinate is
\begin{equation}
\begin{aligned}
\tau(r)
&=T\int_0^r \frac{2}{\mathfrak{a}^4}s^3e^{-s^2/\mathfrak{a}^2}\,ds \\
&= T\[1-e^{-r^2/\mathfrak{a}^2}\(1+\frac{r^2}{\mathfrak{a}^2}\)\].
\end{aligned}
\end{equation}

To invert it for $r(\tau)$, we can use the $-1$ branch of the Lambert $W$ function, obtaining for $\tau\in\[0,T\)$
\begin{equation}
r(\tau)=\mathfrak{a}\sqrt{-1-W_{-1}\(\frac{\tau/T-1}{e}\)}.
\end{equation}

\image{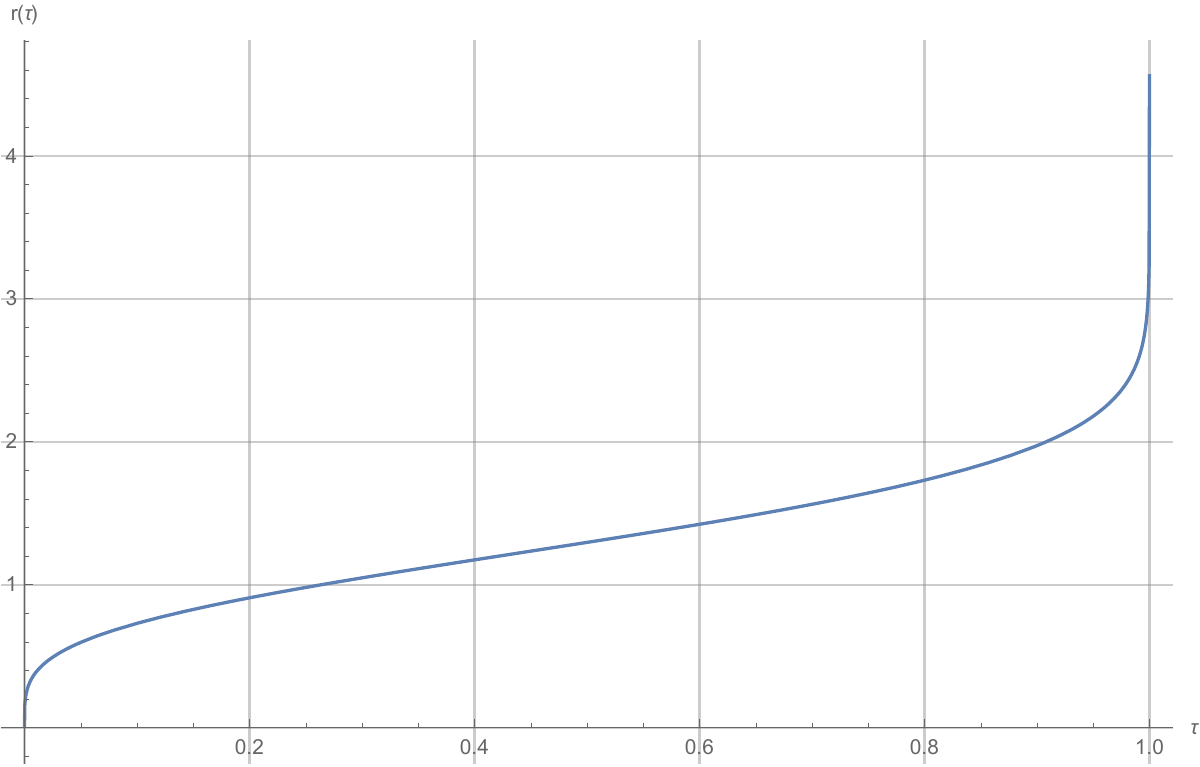}{0.48}{Plot of $r(\tau)$ for the harmonic oscillator potential, with $\tau$ expressed in $T$-units. The solution has a ``Big Rip'' singularity at $\tau\to T$.}

As seen in Figure~\ref{harmonic}, there is a ``Big Rip'' singularity at $\tau\to T$,
\begin{equation}
\lim_{\tau\to T}r(\tau)=\infty.
\end{equation}

In other words, the intrinsic time is finite, $\tau\in\(0,T\)$.
This is a general property of all normalizable solutions.

From~\eqref{eq:dr-of-dtau}, we can compute the Hubble parameter
\begin{equation}
\frac{\dot r}{r}=\frac{1}{T I(r)r}
=\frac{\mathfrak{a}^4 e^{r^2/\mathfrak{a}^2}}{2T r^4}.
\end{equation}

From~\eqref{eq:I-of-r-harm},
\begin{equation}
\label{eq:I-of-r-harm-prime}
I'(r)=I(r)\(\frac{3}{r}-\frac{2r}{\mathfrak{a}^2}\)
\end{equation}
and by differentiating $\dot r$ once more and applying~\eqref{eq:I-of-r-harm-prime},
\begin{equation}
\begin{aligned}
\frac{\ddot r}{r}
&\stackrel{\eqref{eq:I-of-r-harm-prime}}{=}-\frac{1}{T^2 I(r)^2}\(\frac{3}{r^2}-\frac{2}{\mathfrak{a}^2}\) \\
&\stackrel{\eqref{eq:I-of-r-harm}}{=}-\frac{\mathfrak{a}^8}{4 T^2}\(\frac{3}{r^2}-\frac{2}{\mathfrak{a}^2}\)\frac{e^{2r^2/\mathfrak{a}^2}}{r^6}.
\end{aligned}
\end{equation}

The shell density reaches its maximum at $r=r_{\mathrm{max}}$ satisfying $I'(r_{\mathrm{max}})=0$, and from equation~\eqref{eq:I-of-r-harm-prime} we get
\begin{equation}
\frac{3}{r_{\mathrm{max}}}-\frac{2r_{\mathrm{max}}}{\mathfrak{a}^2}=0,
\end{equation}
which gives
\begin{equation}
r_{\mathrm{max}}=\sqrt{\frac{3}{2}}\mathfrak{a}.
\end{equation}

Then, the expansion of this ``harmonic universe'' is decelerating before $\tau(r_{\mathrm{max}})$ and accelerating after.

\subsection{Example: Free Hamiltonian}
\label{s:emergent-cosmology-free}

In fact, we get a curved expanding universe even in the case when $V(r)=0$.
Equation~\eqref{eq:radial-eigenvalue-eq} becomes
\begin{equation}
\Psi''(r)+\frac{3}{r}\Psi'(r)=0.
\end{equation}

Regular solutions at $r=0$ are constant:
\begin{equation}
\Psi(r)=C.
\end{equation}

Equation~\eqref{eq:I-of-r} becomes
\begin{equation}
\label{eq:I-of-r-free}
I(r)=2\pi^2\abs{C}^2r^3,
\end{equation}

From equation~\eqref{eq:tau-of-r} we get the intrinsic time coordinate
\begin{equation}
\tau(r)=\frac{\pi^2T \abs{C}^2}{2}r^4,
\end{equation}
therefore
\begin{equation}
r(\tau)=\sqrt[4]{\frac{2\tau}{\pi^2 T\abs{C}^2}}.
\end{equation}

\image{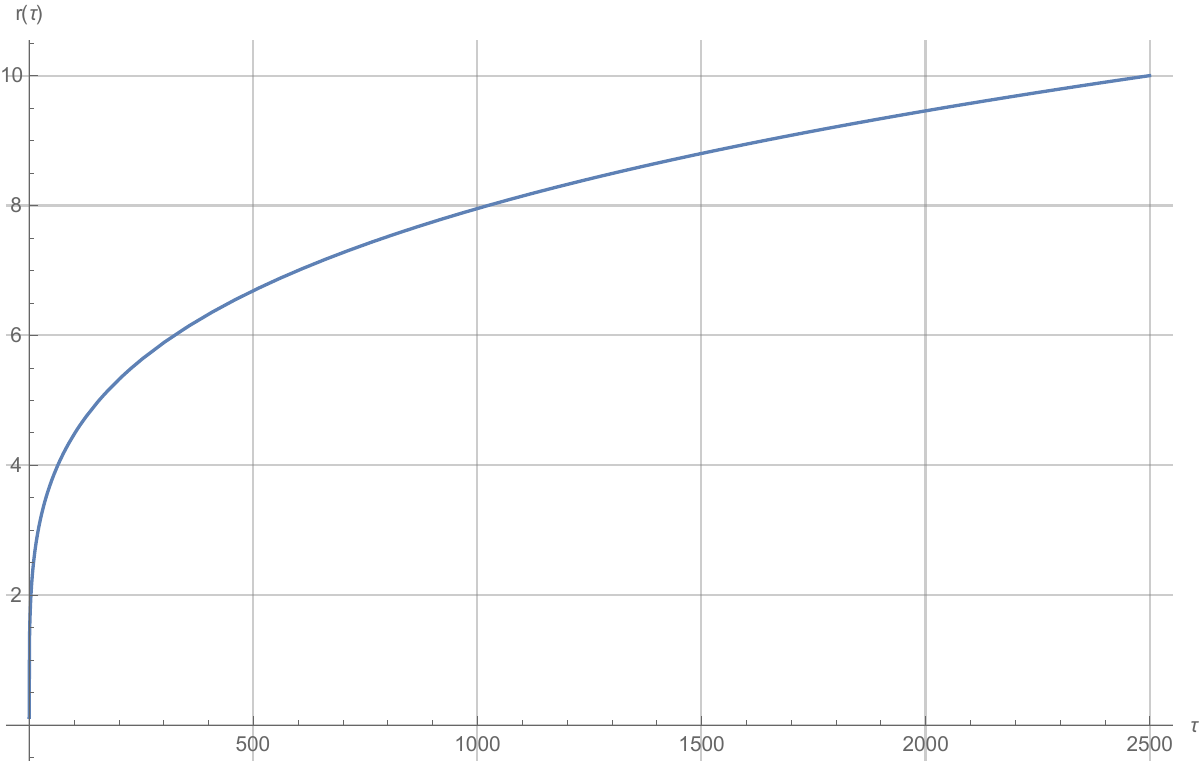}{0.48}{Plot of $r(\tau)$ for the free potential.}

From equations~\eqref{eq:dot-r} and~\eqref{eq:I-of-r-free}, we get
\begin{equation}
\frac{\dot r}{r}=\frac{1}{2\pi^2 T\abs{C}^2 r^4}.
\end{equation}

Also,
\begin{equation}
\frac{\ddot r}{r}=-\frac{3}{4\pi^4 T^2\abs{C}^4 r^8}.
\end{equation}

Since $\ddot r<0$ for all values, the zero-energy free solution gives an expanding universe with decelerated expansion.

While this solution is constant on the Euclidean space, it is an expanding cosmology that emerged merely due to the choice of the spherically symmetric intrinsic time observable, which results in the effective radial potential $V_{\mathrm{eff}}(r)$ from equation~\eqref{eq:potential-effective}.

\subsection{Determining \texorpdfstring{$\Psi$}{Psi} and \texorpdfstring{$V$}{V} from the FLRW geometry}
\label{s:emergent-cosmology-inverse}

The previous examples, which were based on simple potentials, describe expanding FLRW cosmologies, but do not fit the actual cosmological data.
However, we can find suitable potentials and stationary solutions for any closed and expanding FLRW geometry, \ie for any positive monotone function $r=r(\tau)$, as long as, for all $\tau>0$,
\begin{equation}
\label{eq:r-invertible}
\dot r(\tau)>0.
\end{equation}

Condition~\eqref{eq:r-invertible} allows the existence of an inverse $\tau(r)$, and since $\dv*{\tau(r)}{r}=1/\dot r\(\tau(r)\)$, using~\eqref{eq:dtau-of-dr},
\begin{equation}
I(r)=\frac{1}{T\dot r\(\tau(r)\)}.
\end{equation}

From this and equation~\eqref{eq:I-of-r},
\begin{equation}
\abs{\Psi(r)}^2=\frac{1}{2\pi^2 T r^3 \dot r\(\tau(r)\)}.
\end{equation}

The simplest choice for $\Psi$ is the real function
\begin{equation}
\Psi(r)=\frac{1}{\sqrt{2\pi^2 T r^3 \dot r\(\tau(r)\)}}.
\end{equation}

We can determine the potential that generates our FLRW geometry from the radial Schr\"odinger equation~\eqref{eq:radial-eigenvalue-eq}, obtaining
\begin{equation}
\label{eq:V-of-Psi}
V(r)=\frac{\hbar^2}{2m}\frac{\Psi''(r)+\frac{3}{r}\Psi'(r)}{\Psi(r)}.
\end{equation}

In terms of $u(r)=r^{3/2}\Psi(r)$, the potential is
\begin{equation}
V(r)=\frac{\hbar^2}{2m}\frac{u''(r)}{u(r)}-\frac{\hbar^2}{2m}\frac{3}{4r^2}.
\end{equation}

\subsection{Recovering the closed \texorpdfstring{$\Lambda$}{Lambda-}CDM model}
\label{s:emergent-cosmology-Lambda-CDM}

The minimal closed $\Lambda$CDM model is usually given in the form~\cite{DeruelleUzan2018RelativityInModernPhysics}
\begin{equation}
\label{eq:hubble-vel-Lambda-CDM-Omega}
\frac{H^2}{H_0^2}
=\Omega_{m0}\(\frac{a_0}{a}\)^3+\Omega_{r0}\(\frac{a_0}{a}\)^4+\Omega_{\Lambda0}+\Omega_{k0}\(\frac{a_0}{a}\)^2,
\end{equation}
where $a_0=a(\tau_0)$ with $\tau_0$ the present time, $H(\tau)=\dot{a}(\tau)/a(\tau)$ the Hubble parameter, $H_0=H(\tau_0)$. The coefficients $\Omega_{\ast 0}$ are
\begin{equation}
\label{eq:LCDM-Omegas}
\begin{aligned}
\Omega_{m0}&=\frac{8\pi G\rho_{m0}}{3H_0^2},&
\Omega_{r0}&=\frac{8\pi G\rho_{r0}}{3H_0^2}, \\
\Omega_{k0}&=-\frac{k c^2}{H_0^2a_0^2},&
\Omega_{\Lambda0}&=\frac{\Lambda c^2}{3H_0^2}, \\
\end{aligned}
\end{equation}
where $\rho_{m0}$ and $\rho_{r0}$ are the current densities of matter and radiation, $\Lambda$ is the cosmological constant, and $k=1$ (for a closed universe). Equation~\eqref{eq:hubble-vel-Lambda-CDM-Omega} can be written as
\begin{equation}
\label{eq:hubble-vel-Lambda-CDM}
\frac{\dot a^2}{a^2}
=\frac{8\pi G}{3}\rho_{r0}a^{-4}+\frac{8\pi G}{3}\rho_{m0}a^{-3}-c^2a^{-2}+\frac{\Lambda c^2}{3}.
\end{equation}

To find the potential $V$ and the stationary solution $\Psi$ giving this model, we apply equation~\eqref{eq:hubble}, obtaining
\begin{equation}
\abs{\Psi(r)}=\frac{1}{\pi r\sqrt[4]{4T^2 \(\frac{8\pi G}{3}\(\rho_{r0}+\rho_{m0}r\)-c^2r^2+\frac{\Lambda c^2}{3}r^4\)}}.
\end{equation}

The simplest solution is the real one,
\begin{equation}
\label{eq:LCDM-Psi}
\Psi(r)=\frac{R(r)}{\pi \sqrt[4]{4T^2}},
\end{equation}
where
\begin{equation}
R(r)=\frac{1}{r\sqrt[4]{\frac{8\pi G}{3}\(\rho_{r0}+\rho_{m0}r\)-c^2 r^2+\frac{\Lambda c^2}{3}r^4}}.
\end{equation}

Equation~\eqref{eq:V-of-Psi} becomes
\begin{equation}
\label{eq:V-of-Psi-LCDM}
V(r)=\frac{\hbar^2}{2m}\frac{R''(r)+\frac{3}{r}R'(r)}{R(r)}.
\end{equation}

Computing $V(r)$ gives a rational function in $r$
\begin{equation}
\label{eq:V-of-Psi-LCDM-b}
    V(r) = - \frac{\hbar^2}{2m}\frac{v_0 + v_1 r + v_2 r^2 + v_3 r^3 + v_4 r^4 + v_5 r^5 + v_6 r^6}
		{16r^2\(\frac{8\pi G}{3}\(\rho_{r0}+\rho_{m0}r\)-c^2 r^2+\frac{\Lambda c^2}{3}r^4\)^2}
\end{equation}
where
\begin{equation}
\label{eq:V-of-Psi-LCDM-c}
\begin{aligned}
v_0 &=\frac{1024}{9}\pi^2 G^2\rho_{r0}^2 \\ 
v_1 &= 256\pi^2 G^2\rho_{r0}\rho_{m0}\\ 
v_2 &= \frac{320}{3}\pi^2 G^2\rho_{m0}^2-128\pi Gc^2\rho_{r0} \\ 
v_3 &= -\frac{256}{3}\pi Gc^2\rho_{m0}\\ 
v_4 & = 12 c^4 + \frac{256}{3}\pi G\Lambda c^2\rho_{r0}\\ 
v_5 & = \frac{160}{3} \pi G \Lambda c^2\rho_{m0}\\ 
v_6 & = -\frac{32}{3} \Lambda c^4. \\ 
\end{aligned}
\end{equation}

This potential, while more complicated than the specification of $\Lambda$CDM from~\eqref{eq:hubble-vel-Lambda-CDM}, is in a sweet spot between the Big-Rip of the harmonic potential and the free potential.

Let us describe its asymptotic behavior.
At the origin, as $r\to 0$, the dominant term in the potential becomes
\begin{equation}
    V(r) \approx -\frac{\hbar^2}{2m}\frac{v_0}{16r^2\(\frac{8\pi G}{3}\rho_{r0}\)^2}=-\frac{\hbar^2}{2m}\frac{1}{r^2}.
\end{equation}

The effective potential $V_{\mathrm{eff}}(r)$ from~\eqref{eq:potential-effective} becomes 
\begin{equation}
    V_{\mathrm{eff}}(r) \approx -\frac{\hbar^2}{2m}\frac{1}{4r^2},
\end{equation}
exactly matching the threshold for the strongly attractive inverse-square potential. This ensures that the wavefunction remains regular and the local probability density is finite, preventing an infinite intrinsic past time.

As $r\to\infty$, the potential decays as
\begin{equation}
    V(r) \sim -\frac{\hbar^2}{2m}\frac{v_6 r^6}{16r^2\(\frac{\Lambda c^2}{3}r^4\)^2}=\frac{\hbar^2}{2m}\frac{6}{\Lambda r^4}
\end{equation}
and the wavefunction as
\begin{equation}
    \Psi(r) \sim \frac{1}{\pi \sqrt[4]{\frac{4}{3}T^2\Lambda c^2}}\frac{1}{r^2}.
\end{equation}

The asymptotic behavior of its norm is, for large $r$ and fixed $R>0$,
\begin{equation}
\int_R^\infty \abs{\Psi(r)}^2 r^3\,\dd r \sim \frac{1}{2\pi^2 T\sqrt{\frac{\Lambda}{3}}c}\int_R^\infty\frac{1}{r} \,\dd r.
\end{equation}

Since this integral diverges logarithmically, $\abs{\Psi(r)}^2$ is not integrable, the system lying exactly on the threshold between bound and free, and the intrinsic time is unbounded toward the future.

\subsection{Comments}
\label{s:emergent-cosmology-further}

The toy model from this section only predicts a closed expanding FLRW cosmology, assuming spherical symmetry in a four-dimensional Euclidean space. The potential can be tweaked to obtain any closed expanding FLRW cosmology, including the $\Lambda$CDM model, without postulating gravity or curvature.

A common characteristic of the toy models based on a spherically symmetric potential in a four-dimensional Euclidean space presented above is that they can never contract. This is due to the fact that the only degree of freedom discussed is the radial coordinate $r$, and $\tau$ is a monotone function of $r$. In reality, matter is much more complex. There are many additional observables, which of course requires a higher-dimensional configuration space for the wavefunction. In a larger configuration space, $\tau$ can take different values for the same radial coordinate $r$, allowing a non-monotonic dependence of $r$ on $\tau$. In particular, such a universe may expand and then contract into a Big Crunch. But then the solution will not be as simple as the above toy models, because the configuration space would be much larger than the four-dimensional Euclidean space, allowing superpositions of different cosmologies.

Another observation is that the emergence of curvature in these FLRW toy models is due not merely to the uniformization of $\tau$, but also to the intrinsic differences between the slices of different $\tau$: the intrinsic metric is different for different values of $\tau$ because it depends on $r$. While the intrinsic metric for each sphere is due solely to Euclidean geometry and it is independent of $\Psi$, in conjunction with different timeless wavefunctions $\Psi$, one obtains different FLRW geometries due to the different metric component $g_{00}$. If we used a cylinder ($r=\text{const.}$), the same approach would result in a flat geometry regardless of the particular timeless wavefunction $\Psi$.
By contrast, the four-dimensional configuration space used here is homogeneous and isotropic in four dimensions, and, for the hypersurfaces of equal $\tau$, both the intrinsic and the extrinsic geometries are ultimately due to the choice of the time observable $\obs{T}$ to be spherically symmetric and uniform.

\section{Toward emergent spacetime curvature?}
\label{s:emergent-curvature}

We have seen that two major results previously obtained from gravity can be obtained without assuming gravity, from the uniformity of the intrinsic time observable, especially if the Hamiltonian is bounded from below~\cite{Stoica2026QuantumTimeWithoutNegativeEnergy}.
The first of these results is the derivation of a timeless Wheeler-DeWitt equation $\obs{H}\kett{\Psi}=0$, achieved in~\cite{Stoica2026QuantumTimeWithoutNegativeEnergy} merely from the positivity of the Hamiltonian, without assuming gravity. 

The second is the toy FLRW cosmology, obtained in Section~\ref{s:emergent-cosmology} from $\tau$-uniformity only. Stationary solutions of the non-relativistic Schr\"odinger equation with a spherically symmetric potential in four dimensions can be used to obtain any closed isotropic and homogeneous expanding cosmological model, including the $\Lambda$CDM model. This is unexpected, as it does not assume curved spacetime, a quantum theory of gravity, or even gravity at all to begin with. Everything is the result of the intrinsic time perspective.
This is only a toy model, but it proves the possibility that spacetime curvature emerges from a quantum theory that does not presuppose gravity, by a suitable choice of the time observable $\obs{T}$.

One may wonder then whether the full four-dimensional spacetime curvature may emerge in the same way. Could general relativity, perhaps in a modified form as indicated by the galactic rotation curves~\cite{Milgrom1983AModificationOfTheNewtonianDynamicsAsAPossibleAlternativeToTheHiddenMassHypothesis,Bekenstein2004RelativisticGravitationTheoryForTheMONDParadigm,FamaeyMcGaugh2012ModifiedNewtonianDynamicsMONDObservationalPhenomenologyAndRelativisticExtensions} emerge in the same way?

\subsection{Four-dimensional curvature}
\label{s:emergent-curvature-four}

In a complex system like our universe, the intrinsic time observable cannot be unique, but not every observable whose spectrum is $\R$ can play this role.
For example, in the Page-Wootters formalism~\cite{PageWootters1983EvolutionWithoutEvolution}, if we allow any observable whose spectrum is $\R$ to play the role of time, we run into the \emph{clock ambiguity problem}: any history can be obtained just by a suitable choice of a clock subsystem~\cite{Albrecht1995TheoryOfEeverythingVsTheoryOfAnything}. This ambiguity was thought to be eliminated by requiring that the clock does not interact with the rest of the universe~\cite{MarlettoVedral2017EvolutionWithoutEvolutionAndWithoutAmbiguities}, but in fact it turns out that it is worse than previously understood: not only can any history be obtained by choosing a different clock subsystem, but also any Hamiltonian of the world, resulting in a complete inability of the measuring devices to be correlated with the measured properties~\cite{Stoica2026TheClockAmbiguityProblemExtendedOrExtinguished}.
The conclusion of the analysis from~\cite{Stoica2026TheClockAmbiguityProblemExtendedOrExtinguished} is that the physical meanings of the observables have to be taken into account, and these meanings are those that impose the necessary restrictions. 
That is, all observables $(\obs{M}_1,\obs{M}_2,\ldots)$ and $(\widehat{q}_1,\widehat{q}_2,\ldots)$ determining the configuration space on which the wavefunction $\Psi(\tau,\M,\q)$ from equation~\eqref{eq:tau-wavefunction-from-t} is defined should have clearly assigned physical meanings, and any intrinsic time observable $\obs{T}$ should depend on them, $\obs{T}=T(\M,\q)$. This restricts the possible ways in which the time observable can be chosen, but some freedom is still required, because the relativity of reference frames requires the existence of additional time observables even in Newtonian physics~\cite{Stoica2026TheClockAmbiguityProblemExtendedOrExtinguished}.

In special relativity, any inertial reference frame has its own time direction, which can be any direction from within the light cone. Since the interior of the light cone has four dimensions, it is possible to choose four timelike vectors as a basis of the Minkowski spacetime. Let $(x_0=c t,x_1,x_2,x_3)$ be coordinates of the Minkowski spacetime, so that $\pdv*{x_0}$ is timelike and the other $\pdv*{x_j}$ are spacelike. Then, four coordinates $(\tau_0,\tau_1,\tau_2,\tau_3)$ defined so that, for some $\varepsilon\in(0,1)$,
\begin{equation}
\label{eq:timelike-coord}
\left\{
\begin{aligned}
\pdv{\tau_0}&=\pdv{x_0},\\
\pdv{\tau_j}&=\pdv{x_0}+\varepsilon \pdv{x_j},\quad j\in\{1,2,3\},\\
\end{aligned}
\right.
\end{equation}
form a coordinate system such that the equal-$\tau_\mu$ hypersurfaces are spacelike, for all $\mu\in\{0,1,2,3\}$.
Figure~\ref{minkowski-two-times} illustrates such ``timelike'' coordinates in an $1+1$ spacetime.
Timelike coordinates like these are possible in general relativity too, at least locally.

\image{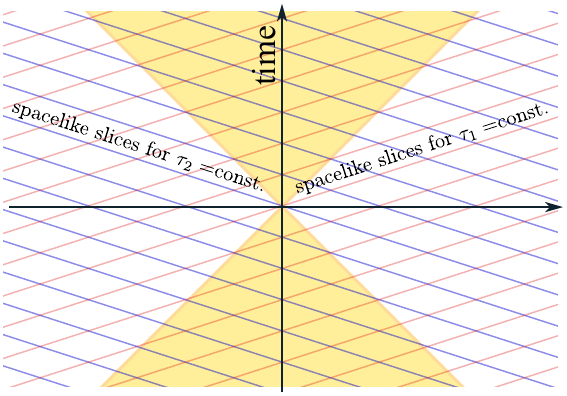}{0.48}{System of timelike coordinates $(\tau_1,\tau_2)$ on a two-dimensional Minkowski spacetime. The red (blue) parallel lines represent equal-$\tau_1$ (equal-$\tau_2$) space slices.}

It is possible to choose all four timelike coordinates to be future-oriented, by replacing some of them from $\tau_\mu$ to $-\tau_\mu$ if necessary, and to pick a common origin for all four.
Then, any physically allowed intrinsic time observable $\tau$ corresponds to a coordinate on the manifold parametrized by $(\tau_0,\tau_1,\tau_2,\tau_3)$. It depends monotonically on all four coordinates, and we can ensure that it increases monotonically.

Returning to our quantum theory that is not assumed to include gravity, the complete set of commuting observables is $\{\obs{T}_0,\obs{T}_1,\obs{T}_2,\obs{T}_3\}\cup\widehat{\M}\cup\widehat{\q}$, so equation~\eqref{eq:tau-wavefunction-from-t} is replaced by
\begin{equation}
\label{eq:tau-wavefunction-from-t-four}
\Psi(\tau_0,\tau_1,\tau_2,\tau_3,\M,\q)=\int_{\R}\psi(\tau_0,\tau_1,\tau_2,\tau_3,\M,\q;t)\dd t.
\end{equation}

\subsection{Toy example: a scalar particle}
\label{s:emergent-curvature-KG}

We will start with the simplest relativistic quantum system, the Klein-Gordon equation for a single scalar particle,
\begin{equation}
\label{eq:klein-gordon}
\[\frac{1}{c^2}\pdv[2]{t}-\nabla^2+\frac{m^2 c^2}{\hbar^2}\]\psi(\x,t)=0.
\end{equation}

We focus on its positive-energy branch
\begin{equation}
\label{eq:klein-gordon-plus}
i\hbar\pdv{\psi}{t}=\obs{H}_+\psi,
\end{equation}
where $\obs{H}_+$ is the Hamiltonian
\begin{equation}
\label{eq:klein-gordon-plus-hamiltonian}
\obs{H}_+=\sqrt{m^2c^4-\hbar^2 c^2\laplacian}.
\end{equation}

The Klein-Gordon equation is too simple to describe even composite systems, and has some flaws that prevent it even from describing a single scalar particle, due to the absence of a direct probabilistic interpretation, but it is relativistically invariant and I will use it as an example.
Probabilities can be understood in extended frameworks like Feshbach-Villars~\cite{FeshbachVillars1958ElementaryRelativisticWaveMechanicsOfSpinZeroAndSpinHalfParticles} or based on Krein spaces~\cite{LangerNajmanTretter2008SpectralTheoryOfTheKleinGordonEquationInKreinSpaces}.
Although the nonrelativistic position observables have no good probabilistic interpretation, we will not need this, but if we did, we could use the Newton-Wigner position operators~\cite{NewtonWigner1949LocalizedStatesForElementarySystems}. Perhaps this would complicate too much the current discussion, intended as a proof of concept.

We will interpret the time-dependent scalar field $\psi(x_1,x_2,x_3;t)$ as a time-independent scalar field $\Psi(x_0,x_1,x_2,x_3)$ defined on the Minkowski spacetime, which in this case is also the configuration space.
Note that, for each $t\in\R$, $\psi(t)$ is a scalar field on $\R^3$, and as usually understood, the space of such fields is independent of $t$.
But since we interpret it as a field on $\R^4$, we will distinguish the spaces of fields on $\R^3$ at different times. In other words, we treat $\psi$ as a section in the bundle $\R_{x_0}\times\R^3\to\R_{x_0}$, where $x_0=c t$.
A way to express this is in Page-Wootters style~\cite{Page1986DensityMatrixOfTheUniverse} as
\begin{equation}
\label{eq:KG-PW}
\kett{\Psi}=\int_\R\ket{x_0}\otimes\ket{\psi\(\frac{x_0}{c}\)}\,\dd x_0.
\end{equation}

This decomposition is similar to $L^2(\R^{m+n})\cong L^2(\R^m)\otimes L^2(\R^n)$ (extended to tempered distributions if needed), and $\ket{x_0}$ needs not be understood as an external clock state.

Based on~\eqref{eq:timelike-coord}, the Klein-Gordon equation~\eqref{eq:klein-gordon} can be expressed in terms of four timelike coordinates $(\tau_0,\tau_1,\tau_2,\tau_3)$ instead of $\x$ and $t$,
\begin{equation}
\label{eq:klein-gordon-four}
\[\pdv[2]{\tau_0}-\frac{1}{\varepsilon^2}\sum_j\(\pdv{\tau_j}-\pdv{\tau_0}\)^2+\frac{m^2 c^2}{\hbar^2}\]\Psi=0.
\end{equation}

In equation~\eqref{eq:klein-gordon} we assumed the Minkowski metric, which is flat. But if we assume $\tau_{\mu}$-uniformity for the four intrinsic time coordinates, this implies curvature in all four dimensions.

Note that we cannot use the usual squared amplitude, since it is not positive definite, but, as mentioned, there are approaches that resolve this problem before the second quantization~\cite{FeshbachVillars1958ElementaryRelativisticWaveMechanicsOfSpinZeroAndSpinHalfParticles,LangerNajmanTretter2008SpectralTheoryOfTheKleinGordonEquationInKreinSpaces}, and we can use one of them.

\subsection{From global to local uniformization}
\label{s:emergent-curvature-local}

In Section~\ref{s:emergent-cosmology} the curvature effect was due to the global squared amplitude density over $\tau$.
But different systems with their own clocks may have different clock rates.
This would require different rates of flow of time at different locations, which may come from the relative local squared amplitudes, rather than from global uniformization.
The example of a scalar particle from~\sref{s:emergent-curvature-KG} illustrates this possibility. However, the density induced by a single scalar field is not sufficient to induce the full Riemann curvature tensor. For example, a four-form applied to a flat spacetime can, at best, induce a conformal transformation resulting in a conformally flat metric~\cite{Penrose1979SingularitiesAndTimeAsymmetry,ONeill1983SemiRiemannianGeometryWithApplicationsToRelativity}.
For the full Riemann curvature tensor, we need more degrees of freedom than simply a single scalar particle.

As an example where the rate of time flow is local, consider the Newtonian limit of general relativity. In this limit, gravity is exclusively due to the metric component $g_{00}$, since
\begin{equation}
\label{eq:newtonian-limit}
\dd s^2=\underbrace{-\(1+\frac{2\Phi\(\x-\x_M\)}{c^2}\)}_{g_{00}}c^2 \dd t^2+\underbrace{\vphantom{\(\frac{\Phi}{\Phi}\)}\dd\x^2,}_{g_{ij}\dd x^i \dd x^j,}
\end{equation}
where $\Phi$ is the Newtonian potential sourced by the mass $m_M$.
So Newtonian gravity can be understood as an effect of the rate of flow of the intrinsic time being different at different locations.

This remains true for the galactic rotation curves~\cite{Milgrom1983AModificationOfTheNewtonianDynamicsAsAPossibleAlternativeToTheHiddenMassHypothesis,Bekenstein2004RelativisticGravitationTheoryForTheMONDParadigm,FamaeyMcGaugh2012ModifiedNewtonianDynamicsMONDObservationalPhenomenologyAndRelativisticExtensions}, which may require a modification of general relativity.
A useful and widely used model has the general form
\begin{equation}
  \label{eq:two-potential-metric}
  \dd s^2 =\underbrace{-\(1+\frac{2\Phi}{c^2}\)}_{g_{00}}c^2\dd t^2+\underbrace{\(1-\frac{2\Phi'}{c^2}\)\dd\mathbf{x}^2.}_{g_{ij}\dd x^i \dd x^j}
\end{equation}
The modified potential $\Phi$ accounts for the galactic rotation curves, while the light deflection is determined by $\Phi+\Phi'$~\cite{FaberVisser2006CombiningRotationCurvesAndGravitationalLensing,CliftonEtal2012ModifiedGravityAndCosmology,DanielEtAl2010LargeScaleStructureAsAProbeOfGravitationalSlip}.

The proposal from this article is that the intrinsic time observables determine the temporal component of the metric $g_{00}$, hence the modified potential $\Phi$, and if there is no preferred time direction, more time observables may also determine the spatial components $g_{ij}$. In particular, for rotational systems like galaxies, both potentials $\Phi$ and $\Phi'$ may be relevant.
But at this point we have merely opened the possibility, there are no quantitative results yet. These require a better understanding of the particular kind of Hamiltonian and intrinsic macroscopic observables, including time.

The fact that curvature may be induced by the amplitudes of $\Psi$ does not show by itself that general relativity emerges from quantum theory simply due to the time observables.
This emergent curvature may be a correction term, maybe the one needed for the galactic rotation curves, or maybe it is a new effect that we may observe.

\section{Intrinsic local clocks and \texorpdfstring{$\tau$}{tau}-uniformity}
\label{s:uniformity-problem}

In many cases, the spectrum of an intrinsic time observable $\obs{T}$ which does not satisfy the uniformity condition~\eqref{eq:prob-tau-uniform} can be reparametrized monotonically, to obtain a uniform intrinsic time observable $\widetilde{\obs{T}}$. We already did this in Section~\ref{s:emergent-cosmology}, whenever we obtained $\tau$ from $r$.
This procedure ensures an effective intrinsic Schr\"odinger equation with respect to $\tau$~\cite{Stoica2026QuantumTimeWithoutNegativeEnergy}.

However, I cannot imagine at this point a straightforward way to ensure the consistency between the uniform intrinsic time observable $\widetilde{\obs{T}}$ and the local clocks we use to define $\obs{T}$ in the first place.
For example, if we use atomic clocks to define $\obs{T}$, it may happen that their periods are not invariant with respect to intrinsic time translations according to the observable $\widetilde{\obs{T}}$.
In this case, we can argue that the right time observable is $\obs{T}$, and not $\widetilde{\obs{T}}$, despite the latter being the uniform one.

We can observe the atomic clocks, and therefore we can observe $\obs{T}$, but can we make intrinsic observations that show us whether $\obs{T}$ is uniform or not?
Normally, one can verify the uniformity of a probability distribution by many repeated trials. However, as an intrinsic part of the universe that we observe, how can we do this for $\obs{T}$, given that we can only experience a single trial of the universe?
Could it be that internal observers experience the squared amplitude $\abs{\Psi(\tau)}^2$ as their subjective flow of time, in a similar way to how we expect to experience probabilities in Everett's interpretation~\cite{Everett1957RelativeStateFormulationOfQuantumMechanics,Wallace2012TheEmergentMultiverseQuantumTheoryEverettInterpretation,Vaidman2012ProbabilityInMWI,Saunders2024FiniteFrequentismExplainsQuantumProbability,Stoica2023TheRelationWavefunction3DSpaceMWILocalBeablesProbabilities}, for example as self-location probabilities~\cite{Stoica2023TheRelationWavefunction3DSpaceMWILocalBeablesProbabilities,Stoica2024ClassicalManyWorldsInterpretation,Stoica2025BornRuleQuantumProbabilityAsClassicalProbability}?

The two rates of flow of the intrinsic time, the one due to self-location in time, and the one shown by internal clocks, must be the same. For example, if
\begin{equation}
\label{eq:time-self-location}
\int_{\text{Monday}}\abs{\Psi(\tau)}^2\dd \tau =10\times\int_{\text{Friday}}\abs{\Psi(\tau)}^2\dd \tau,
\end{equation}
we would find Mondays to be ten times more likely than Fridays, and we would experience this difference as Mondays seeming to last ten times longer than Fridays.
However, what we can notice in reality is that clock readings change on Fridays as much as on Mondays, and our hearts beat more or less the same number of times in the two weekdays.
Since we cannot access $\abs{\Psi(\tau)}^2$ nor can we find observable records of it, there seems to be no physical basis to experience Monday as ten times longer than Friday. The two weekdays should take the same proportion of our time. Then, what correlates the two intrinsic time-flow rates, the one due to observables and the one due to the squared amplitude?

All we can do is to observe the state of the universe at a given intrinsic time. All we can know about past events are the traces they left in the present state, in the form of records. We can read the records of an experiment: how it was prepared, and what result we obtained. We can do many experiments and keep the records. In fact, all that we know, including how often we experienced being Monday \textit{vs.} Friday, is recorded in the intrinsic present time.
These records may indicate that atomic clocks run uniformly with respect to $\obs{T}$, but how can they show whether $\obs{T}$ is uniform as in equation~\eqref{eq:prob-tau-uniform}? This seems to amount to measuring $\abs{\Psi(\tau)}^2$, and we cannot know $\abs{\Psi(\tau)}^2$ from intrinsic observations at some intrinsic time $\tau$, unless there is a reason for the records to encode it.

We can illustrate this problem with the cosmological models from Section~\ref{s:emergent-cosmology}. Based on the intrinsic records, all of these models appear to be identical. In fact, we cannot even talk about an internal perspective when the only parameter that differs from state to state is the radial coordinate $r$, but suppose that there are sufficiently many degrees of freedom for observers to exist. Even so, if all they can know of their present state is $r$, given that there is no way for them to know the squared amplitude $\abs{\Psi(\tau)}^2$, they will be unable to distinguish which of the infinitely many possible cosmologies they inhabit. 
Their intrinsic time observable would be determined by the intrinsic records alone, and not by the squared amplitude $\abs{\Psi(\tau)}^2$.
Therefore, for these models to be different, there must be more than the four-dimensional configuration space, there must be other observables that correlate with the squared amplitude in the right way.

In our toy cosmological models we used a simplification. In the real world, the configuration space for the wavefunction has many more dimensions than only four, perhaps infinitely many, labeled by $\M$ and $\q$. The models from Section~\ref{s:emergent-cosmology} left outside an immense number of degrees of freedom, which normally include the systems that we can use to measure the intrinsic time.
And these degrees of freedom of the configuration space allow an immense range of possibilities for the wavefunction, which results in an immense number of possible values for the observables.
This does not answer the question of how the two rates are correlated, but if there is an answer, this may be the place to look for it.

\section{Conclusions}
\label{s:conclusions}

If the squared amplitude $\abs{\Psi(\tau)}^2$ is relevant for the rate of the flow of time, this implies a new way for curvature to emerge, as discussed in Sections~\ref{s:emergent-cosmology} and~\ref{s:emergent-curvature}. We have seen in Section~\ref{s:emergent-cosmology} that, if we take seriously the uniformity of the time observables, assuming spherical symmetry in four dimensions, we can obtain FLRW cosmologies, including the $\Lambda$CDM model.
The FLRW curvature emerges without assuming gravity.
As seen in Section~\ref{s:emergent-curvature}, this suggests a way by which the four-dimensional spacetime curvature emerges even in a quantum theory in which gravity is not explicitly present.
This suggests a question,
\begin{question}
\label{question:emergent-curvature}
Does gravity emerge in quantum theories that do not contain it explicitly, as an effect of the intrinsic time observables?
Does an additional curvature effect emerge? Can it explain anomalous gravitational effects like the galactic rotation curves?
\end{question}

Section~\ref{s:emergent-curvature} explored a way by which this may work, but it does not provide a quantitative answer.
The present article merely opens a door to a new space of possibilities in which to look for an answer, by suggesting a proof of concept that curvature can emerge as an effect of intrinsic time.

However, the discussion from Section~\ref{s:uniformity-problem} showed that there is no obvious explicit mechanism by which an intrinsic observer can measure the uniform intrinsic time flow compared to other options obtained by any other monotonic reparametrization.
This leaves us with an open question.

\begin{question}
\label{question:why-amplitude}
What explains the correlation between the squared amplitude $\abs{\Psi(\tau)}^2$ and the rate of flow of the intrinsic time observed and recorded from within the world?
\end{question}

This question has a counterpart in Everett's interpretation: ``what correlates the squared amplitude with the outcome frequencies of repeated measurements, observed intrinsically to follow the Born rule?''. But the Everettian question and the time-flow-rate question are, as they appear at this point, different, and their answers seem to require different strategies.

Finally, since the rate of intrinsic time induces curvature,
\begin{question}
\label{question:how-curvature}
If the squared amplitudes that induce spacetime curvature are also encoded intrinsically, how do we know that curvature is induced by the amplitudes, and not an intrinsic feature of the theory?
\end{question}

A more mature version of the present proposal should include at least:
\begin{enumerate}
	\item 
A mechanism for the squared amplitudes to induce both spacetime curvature and the intrinsic records of spacetime curvature.
	\item 
Rigorous conditions for the emergent curvature to be the curvature from general relativity, modified gravity, and/or correction curvature terms.
\end{enumerate}


\end{document}